\documentclass[12pt]{article}

\usepackage{epsfig,amsfonts}

\newcommand{\betac}{\beta_\mathrm{c}}

\begin{document}

\title{
Antiferromagnetic 4-$d$ O(4) Model
}
\author{Isabel~Campos$^{a}$,
Luis~A.~Fern\'andez$^{b}$ and
Alfonso~Taranc\'on$^{a}$}
\bigskip
\maketitle

\begin{center}
{\it a)  Departamento de F\'{\i}sica Te\'orica, Facultad de Ciencias,\\
Universidad de Zaragoza, 50009 Zaragoza, Spain \\
\small e-mail: \tt isabel, tarancon@sol.unizar.es} \\
{\it b) Departamento de F\'{\i}sica Te\'orica I, Facultad de Ciencias
F\'{\i}sicas,\\
Universidad Complutense de Madrid, 28040 Madrid, Spain \\
\small e-mail: \tt laf@lattice.fis.ucm.es} \\
\end{center}
\bigskip
\begin{abstract}

We study the phase diagram of the four dimensional O(4) model with
first ($\beta_1$) and second ($\beta_2$) neighbor couplings,
specially in the $\beta_2 <0$ region, where we find a line of
transitions which seems to be second order.
We also compute the critical
exponents on this line at the point $\beta_1 =0$ ($\rm{F}_4$ lattice) by
Finite Size Scaling techniques up to a lattice size of $24$, being these
exponents different from the Mean Field ones.
\end{abstract}

\newpage

\section{Introduction}

The action for the electroweak sector of the Standard Model has
SU(2)$\times$U(1) symmetry. If we consider the SU(2) part and take
the limit in which gauge degrees of freedom are frozen, the resulting
action is the O(4) non linear $\sigma$ model, which has been extensively
studied because its Spontaneous Symmetry Breaking pattern is related
to the one exhibited by SU(2) in four dimensions.

The regularized version of the O(4) $\sigma$ model on the lattice
leads to an interacting continuum limit for $d < 4$ \cite{BRY}, while for
$d > 4$ the theory is described by free bosonic fields \cite{AIZ,FRO}.

At the upper critical dimension, $d = 4$, deviations from Mean Field
Theory (MFT) are expected. The MFT predictions for the scaling of
thermodynamic quantities are corrected by multiplicative logarithmic
terms \cite{LOG,WEG}. 

Perturbatively the infrared fixed point of the 
Callan-Symanzik function $\beta(g)$ moves to the origin as the
dimension becomes four \cite{LOG}, also the fixed point is now a double
zero (in contrast with the $d < 4$ case) which is responsible for
the occurrence of such logarithmic corrections.

The existence of these corrections imply the triviality of the
theory \cite{LUS}. Triviality seems to persist when gauge
fields are included \cite{CALL,U1H}.

The common feature of all these approaches to the so-called triviality
problem \cite{CALL}, is that the self-interactions of the scalar field
in the broken phase are weak, and they can be reasonably studied
within the context of perturbation theory.
It is generally believed that the perturbative and the strong
coupling regime belong to the same universality class. However, for
the non-perturbative strong coupling regime a rigorous proof is still
lacking and we have to rely on numerical simulations \cite{LANG,LAT}

The existence of
a strongly interacting Higgs sector, with a complicated dynamics,
rendering useless perturbation theory, is a possibility
not to be discarded a priori. A large amount of work
have been done actually in order to know 
whether or not non-perturbative effects could change 
the physics of the electro-weak symmetry breaking sector 
(for a review see \cite{MONT}).
In this sense, concerning the universality class of the RP$^{N-1}$ models,
the role of non-perturbative effects needs to be clarified 
\cite{RP22}.

Antiferromagnetism (AF) has been considered in a great variety of
models in order to find properties not present in the purely
ferromagnetic (FM) systems \cite{RP2,RP24}
In the context of High $T_{\rm c}$ Superconductivity,
AF seems to play an essential role. The transition from paramagnetic
to non purely FM ordered phases has been studied in two dimensional
models \cite{PLUM,KAWA,MORI,ISING2d}.

In four dimensions, in diluted systems recently new critical exponents
have been obtained \cite{PAJJ}.
Also in $d=4$ a previous study of the AF Ising model \cite{ISING} shows
the existence of an AF phase non trivially equivalent to the standard
FM one. However no new critical behavior was evidenced in this work.

Also in four dimensions, competing interactions have been considered
in order to study the multicritical point of the Yukawa models. At this
multicritical point four phases meet (FM, AF, Ferrimagnetic and PM). The
question of whether or not it would be possible to define a non-trivial
continuum limit at this point remains still an open problem 
\cite{BOCK,BOCK2,EBI,JL}.

It is not clear the role that AF can play in the
formulation of QFT, nevertheless it is worthwhile a careful study of
this kind of models since they are known to have very rich phase
diagrams, and presumably new universality classes could appear in which
alternative formulations of continuum QFT should be possible. However,
when defining a theory with AF couplings one has to be aware of 
the fact that higher
order derivatives tends to violate reflection positivity
\cite{GALLA,POLO}. A possibility is to perform
an appropriate tuning of the couplings in order to cancel the 
contributions coming from unphysical (negative norm) states.

The inclusion of gauge fields can change the situation \cite{POLO},
but it is worthwhile as a probe
to see what happens in this limit when negative couplings are included,
postponing for a future study the effect of gauge fields.

In this work we study how the existence of opposite couplings
influence the vacuum of the theory, specifically, whether or not
the Ground state ($\Omega$) is frustrated 
(the energy cannot be minimized simultaneously
for all couplings) or even disordered (non zero vacuum entropy).

\section{The Model}

Our starting point is the non-linear $\sigma$ model, with action:

\begin{equation}
S_{\sigma} = - \beta \sum_{\bf{r},\mu} 
        {\bf \Phi}_{\bf{r}} {\bf \Phi}_{\bf{r}+\hat{\bf{\mu}}}\ . 
\end{equation} 

Where ${\bf \Phi}$ is a 4-component vector with fixed modulus 
${\bf \Phi}_{\bf{r}}\cdot{\bf \Phi}_{\bf{r}}=1$.

The naive way to introduce AF in the non-linear $\sigma$ model is to consider
a negative coupling. In this case the state with minimal energy for large
$\beta$ is a staggered vacuum. On a hypercubic lattice, if we denote the
coordinates of site $\bf{r}$ as $(r_x,r_y,r_z,r_t)$, making the 
transformation
\begin{equation}
{\bf \Phi}_{\bf{r}}\to
        (-1)^{r_x+r_y+r_z+r_t}{\bf \Phi}_{\bf{r}}\ ,
\label{MAPPING}
\end{equation}
the system with negative $\beta$ is mapped onto the positive $\beta$
one, both regions being exactly equivalent.

Therefore to consider true AF we must take into account either different
geometries or more couplings, in order to break the symmetry under the
transformation
(\ref{MAPPING}). In four dimensions the simplest
option is to add more couplings, we have chosen to add a coupling between
points at a distance of $\sqrt{2}$ lattice units.

Following this we will consider a system of spins
\{$\bf{\Phi}_{\bf{r}}$\} taking values in the hyper-sphere 
${\rm S}^3 \subset \mathbf{R}^4$ and placed in the nodes of a cubic lattice.  
The interaction is defined by the action
\begin{equation}
S = - \beta_1 \sum_{\bf{r},\mu} 
        {\bf \Phi}_{\bf{r}} {\bf \Phi}_{\bf{r}+\hat{\bf{\mu}}} 
    - \beta_2 \sum_{\bf{r},\mu<\nu} 
        {\bf \Phi}_{\bf{r}} {\bf \Phi}_{\bf{r}+\hat{\bf{\mu}} + 
 \hat{\bf{\nu}}}\ ,
\label{ACCION}
\end{equation}

The transformation (\ref{MAPPING}) maps the semi-plane 
$\beta_1 >0$ onto the $\beta_1 <0$, and therefore
only the region with $\beta_1 \geq 0$ will be considered. On the line
$\beta_1 =0$ the system decouples in two $\rm{F}_4$ independent sublattices.

When $\beta_2 =0$ the model is known to present a continuous transition
between a disordered phase, where O(4) symmetry is exact, to an ordered
phase where the O(4) symmetry is spontaneously broken to
O(3).
This transition is second order, being the critical
exponents those of MFT: $\alpha=0$, $\nu=0.5$, $\beta=0.5$,
$\eta=0$ and $\gamma =1$ up to logarithmic corrections.
The critical coupling for this case can be studied analytically by an expansion
in powers of the coordination number ($q=2d$), being 
$\beta^{\rm c} = 0.6055 + O(q^{-2d})$ \cite{FISH}.

From a Mean Field analysis, we observe that
for $\beta_2 > 0$ the behavior of the system will not change 
qualitatively from the $\beta_2 =0$ case but with higher coordination number. 
In fact, taking into account that the energy (for non-frustrated systems)
is approximately proportional to the coordination 
number, there will be a transition phase line whose approximate equation
is
\begin{equation}
\beta_1^{\rm c} + q\beta_2^{\rm c} = \beta^{\rm c}\ ,
\label{RECTA}
\end{equation}
where $q$ is the quotient between the number of second and first
neighbors. 
This line can be thought as a prolongation of the critical point
at $\beta_2 =0$ so the transitions on this line are expected
to be second order with MFT exponents.
This is also the behavior of the two couplings Ising model in this
region \cite{ISING}.

When $\beta_2 <0$, the presence of two couplings with opposite sign makes 
frustration to appear, and very different vacua are possible.

\section{Observables and order parameters}

We define the energy associated to each coupling:
\begin{equation}
E_1 \equiv \frac{\partial \log Z}{\partial \beta_1} =
 \sum_{\bf{r},\mu} \bf{\Phi}_{\bf{r}}\cdot \bf{\Phi}_{\bf{r} +
 \hat{\bf{\mu}}}\ ,
\end{equation}
\begin{equation}
E_2 \equiv \frac{\partial \log Z}{\partial \beta_2} =
 \sum_{\bf{r},\mu<\nu} \bf{\Phi}_{\bf{r}}\cdot
  \bf{\Phi}_{\bf{r} + \hat{\bf{\mu}} + \hat{\bf{\nu}}}\ .
\label{ENERGIAS}
\end{equation}

In terms of these energies, the action
reads
\begin{equation}
S = -\beta_1 E_1 -\beta_2 E_2\ .
\end{equation}

It is useful to define the energies per bound as
\begin{equation}
e_1 = \frac{1}{4V} E_1,\ e_2 = \frac{1}{12V} E_2\ ,
\end{equation}
where $V=L^4$ is the lattice volume. With this normalization $e_1$ , $e_2$
belong to the interval $[-1,1]$.

We have computed the configurations which minimize
the energy for several asymptotic values of the parameters. We have only
considered configurations with periodicity two. More complex structures have
not been observed in our simulations.

Considering only the $\beta_1\ge 0$ case, 
we have found the following regions:
\begin{enumerate}

\item{} Paramagnetic (PM) phase or disordered phase, for small absolute
values of $\beta_1,\beta_2$.    

\item{} Ferromagnetic (FM) phase. It appears when $\beta_1+6\beta_2$
is large and positive.

When the fluctuations go to zero, the vacuum takes the form
$\bf{\Phi}_{\bf{r}}= \bf{v}$, where $\bf v$ is an
arbitrary element of the hyper-sphere. 

Concerning the definition of the order parameter let us remark that
because of tunneling phenomena in finite lattice we are forced to use
pseudo-order parameters for practical purposes. Such quantities behave
as true order parameters only in the thermodynamical limit. In the FM
phase, we define the standard (normalized) magnetization as

\begin{equation}
{\bf M}_{\rm F} =\frac{1}{V} \sum_{\bf{r}} {\bf \Phi}_{\bf{r}}\ ,
\end{equation}
and we use as pseudo-order parameter the square root of the norm of
the magnetization vector
\begin{equation}
M_{\rm F} = \langle \sqrt{{\bf M}_{\rm F}^2} \, \rangle\ .
\end{equation}
This quantity has the drawback of being non-zero in the symmetric phase
but it presents corrections to the bulk behavior order $1/\sqrt{V}$.

\item{} Hyper-Plane Antiferromagnetic phase (HPAF). It corresponds
to large $\beta_1$, with $\beta_2$ in a narrow interval 
($[-\beta_1/2,-\beta_1/6]$ in the Mean Field approximation).
In this region the vacuum correspond to spins aligned in three
directions but anti-aligned in the fourth ($\mu$).

In absence of fluctuations the associated vacuum would be
${\bf \Phi}_{\bf{r}} = (-1)^{r_{\mu}} \bf{v}$, 
where $\mu$ can be any direction, and $\bf{v}$ any vector on S$^4$.

We  define an {\em ad hoc} order parameter for this phase as 
\begin{equation}
{\bf M}_{\rm{HPAF},\mu} = \frac{1}{V}
\sum_{\bf{r}}(-1)^{r_{\mu}} {\bf \Phi}_{\bf{r}}\ .
\end{equation}

${\bf M}_{\rm{HPAF},\mu}$ will be different from
zero only in the HPAF phase, where the system becomes antiferromagnetic on the
$\mu$ direction. From the four order parameters (one for every possible value
of $\mu$) only one of them will be different from zero in the HPAF phase.
So, we define as the pseudo order parameter:
\begin{equation}
M_{\rm HPAF} = \sqrt{ \sum_{\mu} {\bf M}_{\rm{HPAF},\mu}^2}\ .
\end{equation} 

\item{} Plane Anti-Ferromagnetic (PAF) phase for $\beta_2$ large and
negative.
In this region the ground state 
is a configuration with spins aligned in two directions and
anti-aligned in the remaining two. It is characterized with by
one of the six combinations of two different directions ($\mu,\nu$),
and an arbitrary spin $\bf v$:
${\bf \Phi}_{\bf{r}} = (-1)^{r_{\mu} + r_{\nu}} \bf{v}$.
For the PAF region we first define 
\begin{equation}
{\bf M}_{\rm{PAF},\mu,\nu} 
= \frac{1}{V} \sum_r (-1)^{r_{\mu} + r_{\nu}} {\bf \Phi}_{\bf{r}}\ ,
\end{equation}
 
and the quantity we measure is
\begin{equation}
M_{\rm PAF} = \sqrt{ \sum_{\mu < \nu} {\bf M}_{\rm{PAF},
(\mu,\nu)}^2 }
\label{MPAF}
\end{equation}

\end{enumerate}

In order to avoid undesirable (frustrating) boundary
effects for ordered phases, we work with even lattice side $L$ as
periodic boundary conditions are imposed.

From this data we can compute the derivatives of any observable with
respect to the couplings as the connected correlation function with
the energies
\begin{equation}
\frac{\partial O}{\partial \beta_j} =
 \langle O E_j \rangle - \langle O \rangle \langle E_j \rangle
\end{equation}

An efficient method to determine $\beta_{\rm c}$ for a second order transition 
is to measure the Binder cumulant \cite{BINDER} for various lattice
size and to locate the cross point in the space of $\beta$.

For O($N$) models $U_L(\beta)$ takes the form \cite{BREZIN}:
\begin{equation}
U_L(\beta) = 1 + 2/N - \frac{\langle ({\bf m}^2)^2 \rangle}
        {{\langle {\bf m}^2 \rangle}^2}
\label{BINDERPAR}
\end{equation} 
where $\bf{m}$ is an order parameter for the transition.

It can be shown \cite{BINDER,BREZIN} that $U_L(0) \rightarrow O(1/V)$ and
$U_L(\infty) \rightarrow 2/N$. 
The slope of $U_L(\beta)$ at $\beta_{\rm c}$ increases with  $L$.

The value of the Binder cumulant is closely related
with the triviality of the theory since the renormalized coupling
(in the massless thermodynamical limit)
at zero momentum can be written as:
\begin{equation}
g_{\rm R} = \lim_{L \rightarrow \infty} g_{\rm R}(L) = 
\lim_{L \rightarrow \infty} (L/\xi_L)^d U_L(\beta_{\rm c})
\label{GR}
\end{equation}
where $\xi_L$ is the correlation length in the size $L$ lattice. 

From this point of view triviality is equivalent to have a vanishing
$g_{\rm R}$ in the thermodynamical limit. In this context it is clear that
we can use the value of $g_{\rm R}$ to classify the universality class.
Out of the upper critical dimension, $L/\xi_L$ is a constant at
$\beta_{\rm c}$ since $\xi \sim L$, and we could use the Binder cumulant
for the same purpose \cite{PAR}. At the upper critical dimension,
$\xi_L$ presents logarithmic corrections and $L/\xi_L$ is no longer a 
constant at $\beta_{\rm c}$. For the FM O(4) model in $d = 4$ (upper critical
dimension) we have perturbatively $L/\xi_L\sim (\ln L)^{-1/4}$ \cite{BRE}. 
In order to have a non trivial theory, the Binder cumulant should
behave as a positive power of $\ln L$, 
but from its definition \cite{BINDER} we see
that $U_L(\beta) \leq 1$. This is just another way of stating the
perturbative triviality of the FM O(4) model. 

\subsection{Symmetries on the $\mathbf{F}_{\bf 4}$ lattice}

In the $\beta_1=0$ case the system decouples in two independent lattices, each
one constituted by the first neighbors of the other. So we consider
two lattices with $\rm{F}_4$ geometry.
There are several reasons to choose the point $\beta_1 = 0$ for a careful
study of the PM-PAF transition. The region with $\beta_1 > 1.5$ evolve
painfully with our local algorithms; For small $\beta_1$ 
we expect very large correlation in MC time because the interaction between 
both sublattices is very small, and the response of one lattice to changes
in the other is very slow. We also remark that the presence of two almost
decoupled lattices is rather unphysical.

We also have the experience from a previous
work for the Ising model \cite{ISING} that the correlation length at
its first order transition is smaller in the $\rm{F}_4$ 
lattice, that means, we can find asymptotic critical behavior in smaller
lattices.

However we should point out that the results in the $\rm{F}_4$
lattice cannot be easily extrapolated to a neighborhood of the
$\beta_1$ axis. Certainly, the geometry of the model is very modified
when $\beta_1 \neq 0$, and perhaps continuity arguments present
problems. Nevertheless, we have run also the case $\beta_1 \sim 0$, and
as occurs in the Ising model we have not found qualitative differences.

In the following when we refer to the size of the lattice $L$ on the
$\rm{F}_4$ lattice we mean a lattice with $L^4/2$ sites. 

We have to find the configurations that maximize $E_2$ in order to define 
appropriate order parameters for the phase transition.

The system has a very complex structure. As starting point we have studied
numerically the vacuum with $\beta_2 \ll 0$. For this 
values we have found in the simulation:

\begin{enumerate}
\item{} The vacuum has periodicity two. 
To check this, we have defined: 
\begin{equation}
{\bf V}_i = \frac{1}{L^d/2^d} \sum_I {\bf \Phi}_{I_i}\ ,
\label{VMAG}
\end{equation}
where $i=0, \dots ,7$ stands for the $i^{th}$ vertex of each $2^4$
hypercube belonging to the $\rm{F}_4$ lattice, and with $I$ we denote the
$2^4$ hypercubes themselves.

From these vectors we can define the 8 magnetizations associated to the
elementary cell,
\begin{equation}
V_i = \langle \sqrt{{\bf V}_i^2} \rangle\ ,
\end{equation}

We have checked that all $V_i$ tends to 1 for the ordered phase in the
thermodynamical limit, so we conclude that the ordered vacua have
periodicity two.

Let us remark for the sake of completeness that all order parameters we have
defined can be written as an appropriate linear combination of the 
${\bf V}_i$.
\item{} In the elementary cell,
$\bf{\Phi}_{\bf{r}+\hat{\bf{\mu}}+ \hat{\bf{\nu}}} =
\bf{\Phi}_{\bf{r}}$ $\forall \mu, \nu$ with $\mu<\nu$. So, in
this section we will restrict the study of the vacuum structure to the
four sites ($i=0,1,2,3$) belonging to the cube in the hyper-plane
$r_t=0$.
\item{} We have measured the energy per bound associated to the
second neighbors coupling. We check that in the thermodynamical
limit $e_2 = -1/3$.
\item{} If we choose the symmetry breaking direction by keeping fix 
one vector, (eg. ${\bf \Phi}_0$) we find:
\begin{equation}
\sum_{i=1}^{3} \left((\bf{\Phi}_0\cdot\bf{\Phi}_i)\bf{\Phi}_0
        -\bf{\Phi}_i\right) = 0 \ ,
\end{equation}

\end{enumerate}

The vacuum structure is not completely fixed by these three conditions
since different symmetry breaking patterns are possible.
For instance, a configuration
${\bf \Phi}_0 = (1,0,0,0)$,  
${\bf \Phi}_1 = (-1/3,\frac{2\sqrt{2}}{3}{\bf v_1})$,
 ${\bf \Phi}_0 =(-1/3,\frac{2\sqrt{2}}{3}{\bf v_2}) $,  
${\bf \Phi}_0 =(-1/3,\frac{2\sqrt{2}}{3}{\bf v_3}) $, 
with {\bf $v_i$} a 3-component
unitary vector with the constraint $\sum_{i \neq j}{\bf v_i v_j} = 0$, 
breaks O(4), but an O(2) symmetry remains (for the different {\bf $v_i$}).

To determine which is the vacuum in presence of fluctuations,
we consider four independent fields in a $2^4$ cell with
periodic boundary conditions.
Let us first consider an O($2$) group.
 We can study the four vectors as a mechanical system of mass-less
links of length unity, rotating in a plane around the same point, whose
extremes are attached with a spring of natural length zero. The
energy for the system is:

\begin{equation}
E = - \sum_{i,j=0, i>j}^{3} \cos(\theta_i - \theta_j)\ .
\label{ENERO2}	
\end{equation}

We consider the fluctuation matrix, $H = \partial E^2 / \partial \theta_i
\partial \theta_j$ in order to find the normal modes. 
The matrix elements of $H$ take the form:

\begin{equation}
H_{i,j} = \delta_{ij} \sum_{k \neq i} \cos(\theta_i - \theta_k) -
                \cos(\theta_i - \theta_j)(1 - \delta_{ij})\ ,
\end{equation}

In the FM case the minimum correspond to $\theta_i=\phi$, 
for all $i$. There is a single zero mode, and a three times degenerated
non-zero mode with eigenvalue $\lambda = -4$. 

For the AF (maximum energy) case, the maximum energy is found, up to
permutations, at $\theta_0=\phi$, $\theta_1 = \phi + \pi$, $\theta_2 =
\phi + \alpha$ and $\theta_3 = \phi + \pi + \alpha$, $\forall \alpha$. 
In addition to
the $\phi$ freedom that corresponds to the global O(2) symmetry, 
there is a degeneration of the vacuum in the $\alpha$ angle
and this zero mode is double $\forall \alpha$.

The  other two eigenvalues are $\lambda_{1,2} = 2(1 \pm \cos\alpha)$,
so, an additional zero mode appears when $\alpha=0$, obtaining
in this case a three fold degenerated zero mode corresponding to:
$\theta_0 = \theta_1 = \theta_2 + \pi = \theta_3 + \pi$. 

The O(4) case is qualitatively similar. We have 12 degrees of
freedom. Of all configurations that minimize the energy, that with a
largest degeneration (9-times) consist of 2 spins aligned and 2
anti-aligned that correspond to a PAF vacuum.
We consider this degeneration as the main difference with the FM sector,
and could be relevant to obtain different critical exponents.

In presence of fluctuations the configurations with largest 
degeneration are favored by phase space considerations, so we expect
that the real vacuum is a PAF one. This statement will be checked
below with Monte Carlo data in the critical region.

\section{Finite Size Scaling analysis}

Our measures of critical exponents 
are based on the FSS ansatz \cite{ITZ,CARDY}. Let be
$\langle O(L,\beta) \rangle$ the mean value of an observable measured
on a size $L$ lattice at a coupling $\beta$. If 
$O(\infty,\beta) \sim \vert \beta - \beta_{\rm c} \vert^{x_O}$, from the FSS
ansatz one readily obtains \cite{CARDY}
\begin{equation}
\langle O(L,\beta) \rangle = L^{x_O/\nu} F_O(L/\xi(\infty,\beta)) +
\ldots \ ,
\label{FSS}
\end{equation}
where $F_O$ is a smooth function and the dots stand for corrections to
scaling terms. 

To obtain $\nu$ we apply equation (\ref{FSS}) to the
operator $\rm{d} \log M_{\rm{PAF}}/\rm{d} \beta$ 
whose related $x$ exponent is
1. As this operator is almost constant in the critical region, 
we just measure at the extrapolated critical point or any 
definition of the apparent critical point in a finite lattice, the
difference being small corrections-to-scaling terms.

For the magnetic critical exponents the situation is more involved
as the slope of the magnetization or the unconnected susceptibility
is very large at the critical point.

We proceed as follows (see refs. \cite{RP2} for other applications
of this method). Let be $\Theta$ any operator
with scaling law $x_{\Theta} = 1$ (for instance
the Binder parameter or a correlation length defined in a finite
lattice divided by $L$).
Applying eq. (\ref{FSS}) to an arbitrary operator, $O$,  and to $\Theta$
we can write
\begin{equation}
\langle O(L,\beta) \rangle = 
        L^{x_O/\nu} f_{O,\Theta}(\langle\Theta(L,\beta)\rangle)+
        \ldots \ .
\end{equation}
Measuring the operator $O$ in a pair of lattices of sizes $L$ and $sL$
at a coupling where the mean value of $\Theta$ is the same, one
readily obtains
\begin{equation}
\left.\frac{\langle O(sL,\beta) \rangle}{\langle O(L,\beta) \rangle}
\right|_{\Theta(L,\beta)=\Theta(sL,\beta)} = s^{x_O / \nu} + \ldots \ .
\label{QUOTIENT}
\end{equation}
The use of the spectral density method (SDM) \cite{FS} avoids an 
exact a priori knowledge of the 
coupling where the mean values of $\Theta$ cross. 
We remark that usually the main source
of statistical error in the measures of magnetic exponents is the
error in the determination of the coupling where to measure. However, 
using eq. (\ref{QUOTIENT}) we can take into account the correlation
between the measures of the observable and the measure of the
coupling where the cross occurs. This allows to reduce
the statistical error in an order of magnitude.

\subsection{FSS at the upper critical dimension: logarithmic corrections}

It is well known that $d = 4$ is the upper critical dimension 
of the FM O(4) model. As we have already pointed out
logarithmic corrections to the Mean Field predictions are expected.
In particular, FSS in its standard formulation breaks at $d = 4$ because
the essential assumption, namely $\xi_L(\beta_{\rm c}) \sim L$ 
is no longer true. In fact, in four dimensions \cite{LUS}:
\begin{equation}
\xi (\infty, t)  \sim  \vert t \vert^{-1/2} 
\vert \ln \vert t \vert \vert^{\frac{1}{4}}   \ .
\end{equation}

The FSS formula for the correlation length was calculated by
Brezin \cite{BRE}. At the critical point one gets:
\begin{equation}
\xi (L, \beta_{\rm c}) \sim L(\ln L)^{1/4} \ .
\end{equation}

It has been suggested \cite{LANG} that the usual FSS statement should
be replaced by the more general one:
\begin{equation}
\frac{O(L, \beta_{\rm c})}{O(\infty, \beta)} =
F_O\left(\frac{\xi(L, \beta_{\rm c})}{\xi(\infty, \beta)}\right)\ .
\end{equation}

When applying the quotient method described above to systems in four
dimensions one has to take into account the logarithmic corrections,
so that the modified formula reads:
\begin{equation}
\left.\frac{\langle O(sL,\beta) \rangle}{\langle O(L,\beta) \rangle}
\right|_{\Theta(L,\beta)=\Theta(sL,\beta)} = 
s^{x_O / \nu}\left(1 + \frac{\ln s}{\ln L}\right)^{1/4} \ .
\label{QUOLOG}
\end{equation}

This point is particularly important when measuring the magnetic critical
exponents because as we have already mentioned, the slope of the 
magnetization and susceptibility are very large, and one has to be very
careful when locating the coupling where to measure.

\section{Numerical Method}

We have simulated the model in a $L^4$ lattice with periodic
boundary conditions. The biggest lattice size has been $L=24$.
For the update we have employed a combination of Heat-Bath and Over-relaxation
algorithms (10 Over-relax sweeps followed by a Heat-Bath sweep).

The dynamic exponent $z$ we obtain is near 1.
Cluster type algorithms are not
expected to improve this $z$ value. In systems with competing
interactions the cluster size average is a great fraction of the whole
system, loosing the efficacy they show for ferromagnetic spin systems.

We have used for the simulations ALPHA processor based machines. The total
computer time employed has been the equivalent of two years of
ALPHA AXP3000. We measure every 10 sweeps and store the individual measures
to extrapolate in a neighborhood of the simulation coupling by using
the SDM.

In the $\rm{F}_4$ case, we have run about $2\times10^5\tau$ for each lattice size, 
being $\tau$ the
largest integrated autocorrelation time measured, that corresponds to
$M_{\rm PAF}$, and ranges from 2.3 measures for $L=6$ to 8.9 for $L=24$.
 We have discarded more than $10^2\tau$ for
thermalization.  The errors have been estimated with the jack-knife
method.

\section{Results and Measures}

\subsection{Phase Diagram}

We have studied the phase diagram of the model using a $L=8$ lattice. We
have done a sweep along the parameter space of several thousands
of iterations, finding the transition lines shown in Figure 1.
The symbols represent the coupling values where a peak in the order parameter
derivative appears.

\begin{figure}[t]
\epsfig{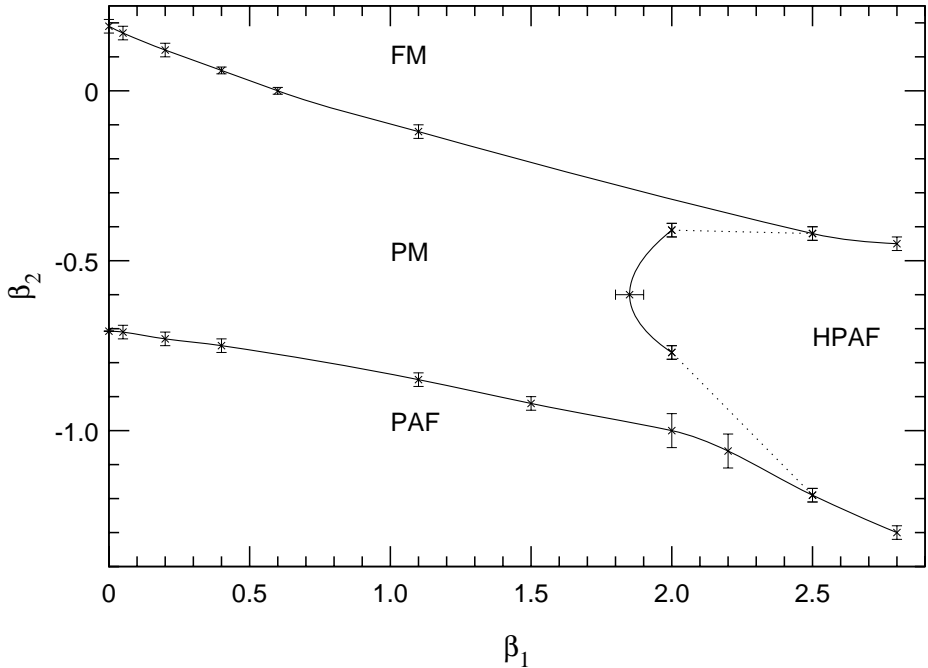}
\caption{Phase diagram obtained from the MC simulation on a $L=8$ lattice} 
\label{PHASE}
\end{figure}

The line FM-PM has a clear second order behavior.
It contains the critical point for the O(4) model with first neighbor
couplings ($\beta_1\approx 0.6$, $\beta_2=0$) with classical exponents
($\nu=0.5$, $\eta=0$).
In the $\beta_1 = 0$ axis, we have computed the critical coupling
($\beta_2^{\rm c} \approx 0.18$) and the critical exponents as a test 
for the method in
the $\rm{F}_4$ lattice. We have also considered the
influence of the logarithmic corrections when computing the exponents.

The lines FM-HPAF, HPAF-PAF and PM-HPAF show clear metastability,
indicating a first order transition.

The regions between the lower dotted line and the PAF transition line,
and between the upper dotted line and the FM transition line, are
disordered up to our numerical precision. We could expect always a PM
region separating the different ordered phases, however, from a MC
simulation it is not possible to give a conclusive answer since the
width of the hypothetical PM region decreases when increasing
$\beta_1$, and for a fixed lattice size there is a practical limit in
the precision of the measures of critical values.

On the line PM-PAF we have found no signs of first order. We have done
hysteresis cycles in several points and no metastability has been observed.
In Figure 2 we plot the energy distribution at the coupling
where a peak in the specific heat appears. There is no evidence of
two-state signal up to $L=24$.

\begin{figure}[!b]
\epsfig{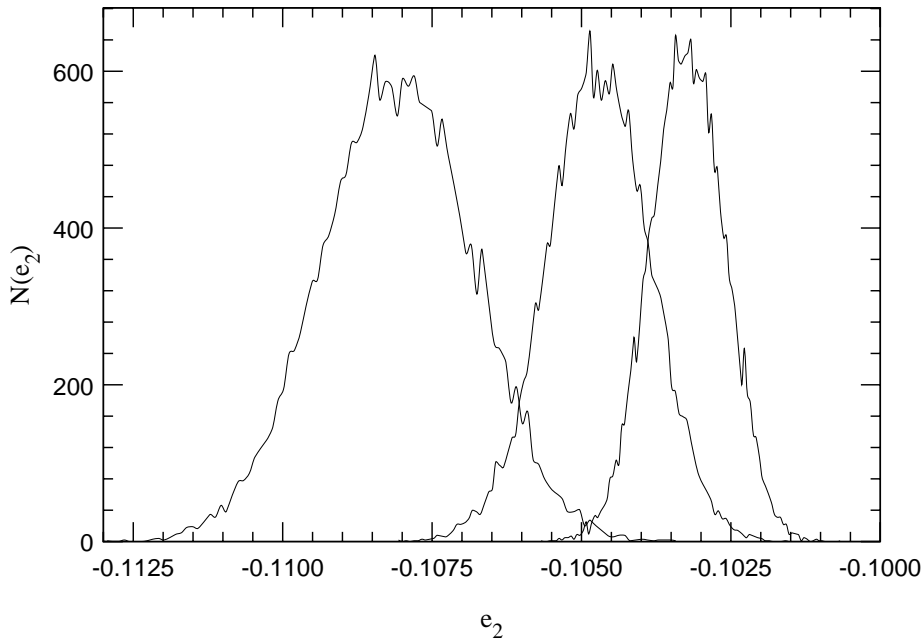}
\caption{Energy distribution for $L$=16,20 and 24 at the peak
of the specific heat on the F$_4$ lattice.}
\label{TODOSH}
\end{figure}

The likely second order behavior of the PM-PAF transition line contrast
with the first order one found in the Ising model with two couplings in
the analogous region \cite{ISING}. This is not surprising because we
are dealing now with a global continuous symmetry. The spontaneous 
symmetry breaking of such symmetries manifest in the appearance of soft
modes or low energy excitations (long wavelength), the Goldstone bosons
in QFT terminology \cite{BELLAC}. 
The role of these soft modes is quite important and
is actually under a a vigorous discussion in the two dimensional case
\cite{ALLES,SEILER}. In general, these low energy modes will perturb
the mechanism of long distance ordering, softening in this way the
phase transitions.

Regarding the differences with the FM case, the most remarkable
feature is the different vacuum structures appearing, specially the very large
degeneration in the PAF transition, in contrast with the single
degeneration of the FM O(4) mode.

As the simpler point for study the properties of the transition,
namely the critical exponents, is the $\rm{F}_4$ limit, most of the MC
work has been done for this case.

\subsection{Results on the F$_{\mathbf 4}$ lattice}

\subsubsection{Results on the FM region}
Firstly, we have checked our method on the FM region of the $\rm{F}_4$
lattice. In Figure 3 the crossing points of the Binder cumulant for
various lattice sizes are displayed. The prediction for the critical
coupling $\beta_{\rm c} \sim 0.1831(1)$ agrees with an earlier study by
Bhanot \cite{BHA2}.

\begin{figure}[!b]
\epsfig{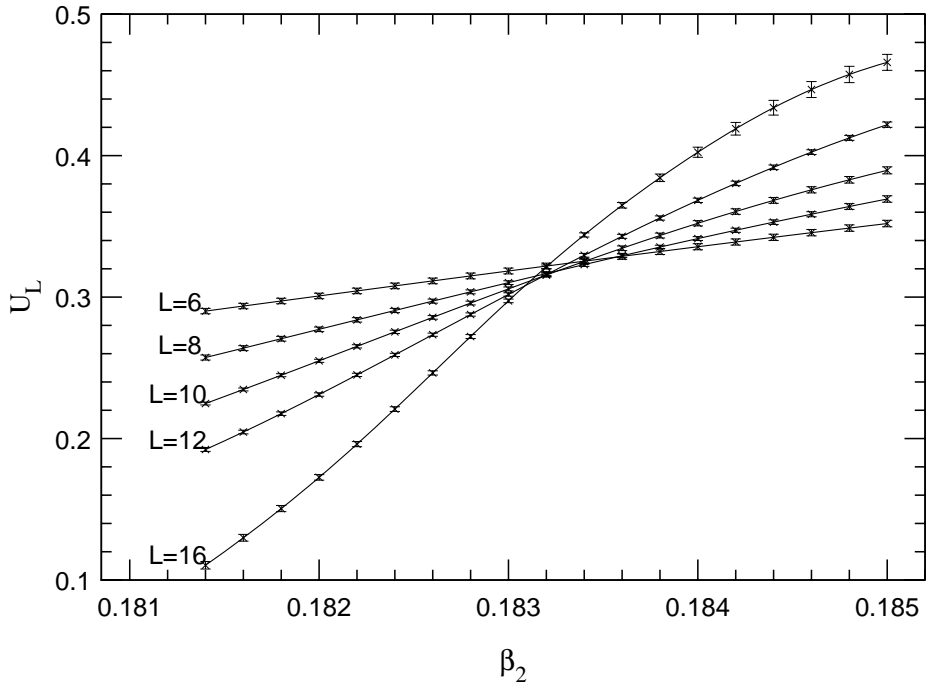}
\caption{Crossing points of the Binder Cumulant for various lattice
sizes on the FM-PM phase transition}
\label{BINDER_F}
\end{figure}

Concerning the measures of critical exponents, we have applied the
quotient method, described in section IV. In table \ref{TABLE_LOG}
we quote the results when logarithmic corrections are included
(formula (\ref{QUOLOG})), and also for sake of comparison, when they
are neglected (formula (\ref{QUOTIENT})). We see how in fact the agreement
of the critical exponents with the MFT predictions is better when the
logarithmic corrections are taken into account.

\begin{table}[h]
{

\begin{center}
{
\begin{tabular}{|c|c|c|c|}\hline
L values  &$8/16$  &$12/16$  &$10/12$  \\ \hline
\multicolumn{4}{|c|}{(without logarithmic corrections)} \\ \hline
$\alpha$/$\nu$                    &0.08(5)    &0.02(2)  &0.13(12) \\
\hline
$\beta$/$\nu$
  &0.92(3)   &0.94(3) &0.87(4)  \\ \hline
$\gamma$/$\nu$
  &2.16(2)  &2.12(2) &2.24(4)  \\ \hline
\multicolumn{4}{|c|}{(with logarithmic corrections)} \\ \hline
$\alpha$/$\nu$                      &0.0   &0.0  &0.03(8) \\
\hline
$\beta$/$\nu$
  &1.04(3)  &1.06(3) &1.04(2)  \\ \hline
$\gamma$/$\nu$
  &1.94(3)  &1.90(4) &1.93(3)  \\ \hline
\end{tabular}
}
\end{center}
}
\caption[a]{Critical exponents for the FM-PM phase transition in the
$\rm{F}_4$ lattice.}
\protect\label{TABLE_LOG}

\end{table}

From now on we will focus on the transition between the PM phase
and the PAF phase on the $\rm{F}_4$ lattice.

\subsubsection{Vacuum symmetries on the PAF region}

We will check using MC data that the ordered vacuum in the critical
region is of type PAF.

Let us define
\begin{equation}
A_{ij}={\bf V}_i\cdot{\bf V}_j\ .
\end{equation}
The leading ordering corresponds to the eigenvector associated to the
maximum eigenvalue of the matrix $A$, that should scale as
$L^{-2\beta/\nu}$ at the critical point.
The scaling law of the biggest eigenvalue agrees with the $\beta/\nu$
value reported in Table \ref{TABLE_EXPO}, and the associated eigenvector
is, within errors, (1,1,-1,-1).

We also have found that the other eigenvalues scale as
$L^{-4}$.  This is the expected behavior if just the O(4) symmetry is
broken, and it remains an O(3) symmetry in the subspace orthogonal to
the O(4) breaking direction.

\subsubsection{Critical Coupling}

To obtain a precise determination of the critical point, $\beta_{\rm c}$,
we have used the data for the Binder parameter (\ref{BINDERPAR}).

In Figure 4 we plot the crossing points of the Binder
cumulants for the simulated lattices sizes. Extrapolations have been done
using SDM from simulations at $\beta_2 = -0.7090$ for
$L=6,8,10,12$ and 16; $\beta_2 =-0.7078$ for $L=20$, and
$\beta_2=-0.7070$ for $L=24$.

\begin{figure}[!t]
\epsfig{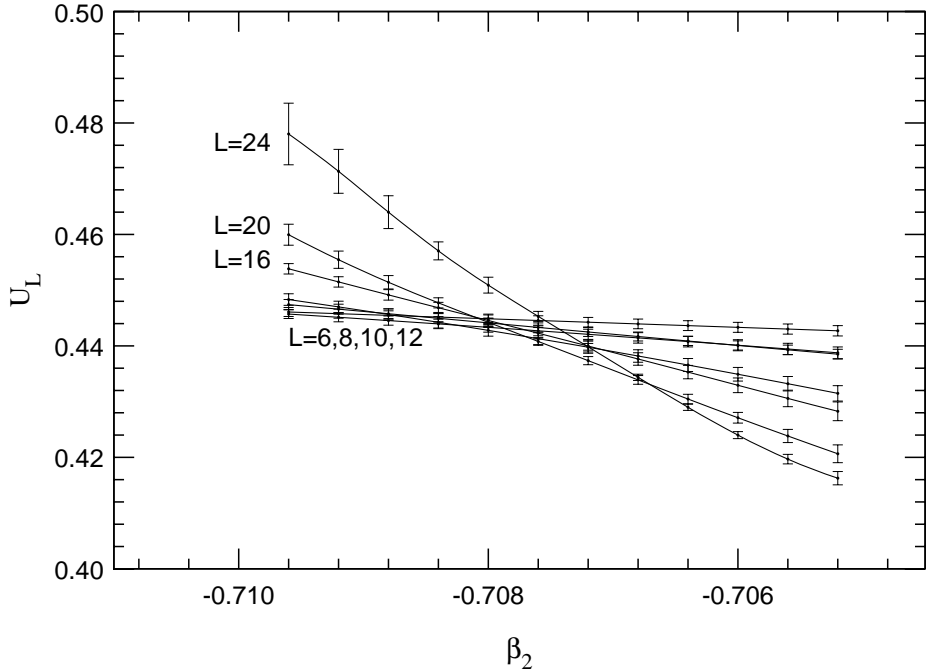}
\caption{Crossing points of   the Binder cumulant for various  lattice
sizes.}
\label{BINDER}
\end{figure}

The shift of the crossing point of the curves  can be explained through the
finite-size confluent corrections. The dependence in the deviation
of the crossing point for $L$ and $sL$ size lattices was estimated by 
Binder \cite{BINDER}
\begin{equation}
\beta_{\rm c}(L,sL) - \beta_{\rm c} 
\sim \frac{1 - s^{-\omega}}{s^{1/\nu} - 1}L^{-\omega - 1/\nu}\ ,
\label{SHIFT}
\end{equation}
where $\omega$ is the universal exponent for the
corrections-to-scaling.

The infinite volume critical point the value
\begin{equation}
\beta_{\rm c}=-0.7065(5)[+2][-2]\ ,
\label{BETAC}
\end{equation}
where the  errors in brackets correspond to the variations in the
extrapolation when we use the values $\omega=0.5$ and $\omega=2$
respectively. In Figure 5 we plot eq. (\ref{SHIFT}) for $\omega=1$.

\begin{figure}[!t]
\epsfig{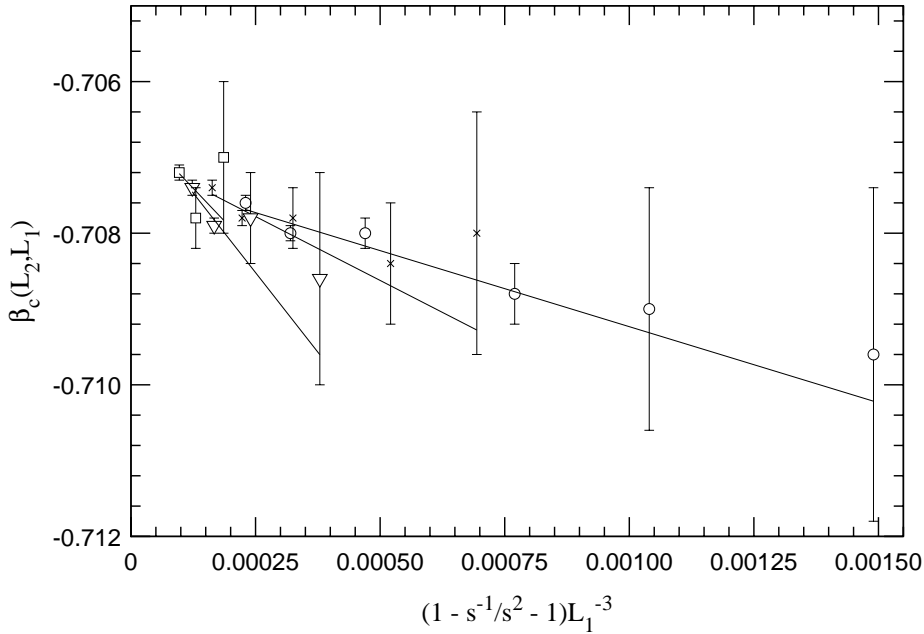}
\caption{Extrapolation to $\betac(\infty)$ for 
$L_1 =6,8,10,12$ (circle,cross,triangle and square symbols respectively).}
\label{BINFI}
\end{figure}

Using the previous value of $\beta_{\rm c}$ we can compute the Binder
cumulant at this point. In table
\ref{TABLE_BINDER} we quote the obtained values. The result points to
that the Binder cumulant stays constant in the critical region. This
result would be compatible with a non zero value of the renormalized
coupling when $L$ increases.

\begin{table}[h]
{
\begin{center}
{
\begin{tabular}{|c|c|c|c|}\hline
Lattice sizes  &$U_L(\beta_{\rm c}(\omega=0.5))$ &$U_L(\beta_{\rm c}
		(\omega=1))$ 
               &$U_L(\beta_{\rm c}(\omega=2))$        \\ \hline
$6$       &0.4435(15) &0.4437(12)   &0.4438(12)           \\   \hline
$8$       &0.4406(15) &0.4409(12)   &0.4413(12)            \\   \hline
$10$      &0.4407(14) &0.4411(16)   &0.4414(14)            \\  \hline
$12$      &0.436(4) &0.437(3)   &0.438(3)            \\  \hline
$16$      &0.435(3) &0.436(3)   &0.437(3)            \\   \hline
$20$      &0.429(5) &0.431(5)   &0.433(5)            \\  \hline
$24$      &0.428(6) &0.430(7)   &0.433(7)            \\  \hline
\end{tabular}
}
\end{center}
}
\caption[a]{ Binder cumulant for various lattices sizes at the 
extrapolated critical point for $\omega = 0.5,1,2$.}
\protect\label{TABLE_BINDER}

\end{table}

Concerning the possibility of having logarithmic corrections in the
determination of the critical coupling, from the numerical point of
view, it is not possible
to discern between the $\omega$ effect, and a logarithmic correction.

\subsubsection{Thermal Critical exponents: $\alpha$, $\nu$}

The critical exponent associated to correlation length can be obtained
from the scaling of:
\begin{equation}
\kappa = \frac{\partial \log{M}}{\partial \beta}\ ,
\end{equation}
where $M$ is an order parameter for the transition, $M_{\rm PAF}$
for our purposes. In the critical region $\kappa \sim L^{1/\nu}$. As
$\kappa$ is a flat function of $\beta$, is not crucial the point where
we actually measure. The results displayed in table \ref{TABLE_EXPO}
have been obtained measuring at the crossing point of the Binder
parameters for lattice sizes $L$ and $2L$ using (\ref{QUOTIENT}).

For measuring $\alpha / \nu$ we study the scaling of the specific heat
\begin{equation}
C = \frac{\partial \langle E_2 \rangle}{\partial \beta_2}\ ,
\end{equation}
We expect that $C$ scales as $A+BL^{\alpha/\nu}$, where $A$ is usually
non-negligible.  In Figure 6 we plot the specific heat
measuring at (\ref{BETAC}), as well as at the peak of the specific
heat, as a function of $L$. We observe a linear behavior for intermediate
lattices. For the largest lattice the slope decreases. The weak first
order behavior~\cite{POTTS} ($\alpha/\nu=1$ for small lattices that
becomes $d$ for large enough sizes) seem hardly compatible with our 
data. If we neglect the $A$ term (what is asymptotically correct), 
and compute the exponent using eq. (\ref{QUOTIENT}) we obtain 
$\alpha/\nu \approx 0.3$ for intermediate lattices that reduces to
$\alpha/\nu=0.15(2)$ for the (20,24) pair.

\begin{figure}[!t]
\epsfig{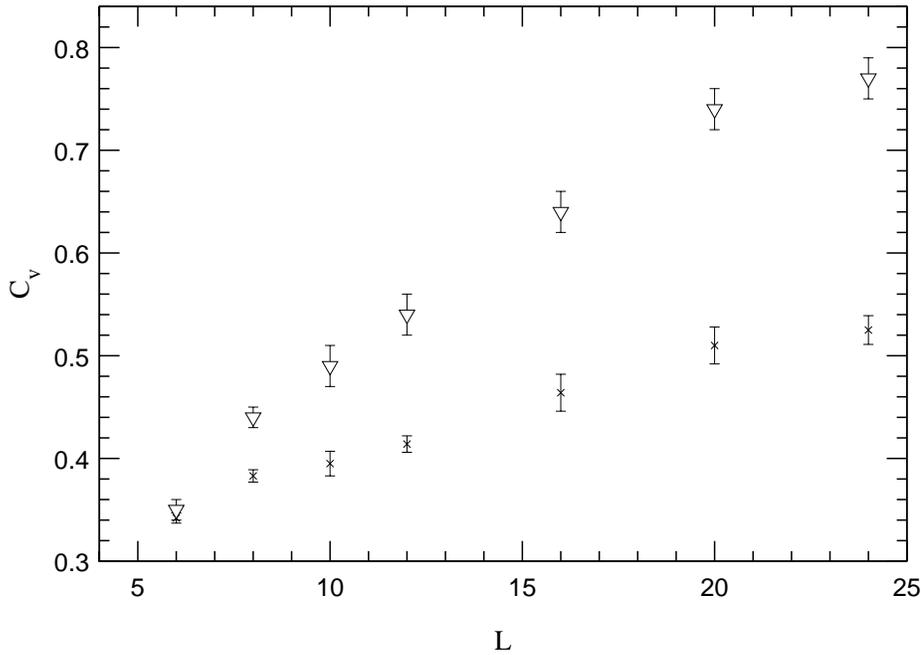}
\caption{Specific heat at the peak (triangle symbols) and at $\beta=-0.7068$
(cross symbols) as a function of the lattice size.}
\label{HEAT}
\end{figure}

However, to give a conclusive answer for the value of $\alpha$ 
statistics on larger lattices are mandatory.

\subsubsection{Magnetic Critical exponents: $\gamma$, $\beta$}

The exponents $\gamma$ and $\beta$ can be obtained respectively from
the scaling of susceptibility and magnetization:
\begin{equation}
\chi \equiv V \langle M^2 \rangle \sim L^{\gamma / \nu} 
\end{equation}
\begin{equation}
M \sim L^{-\beta / \nu}
\end{equation}

Where $M$ is an order parameter for the phase transition.  In Figure 7
upper part, we plot the quotient between $M_{\rm PAF}$
for lattices $L$ and $2L$ as a function of the quotient between the
Binder cumulants for both lattice sizes.

\begin{figure}[t]
\epsfig{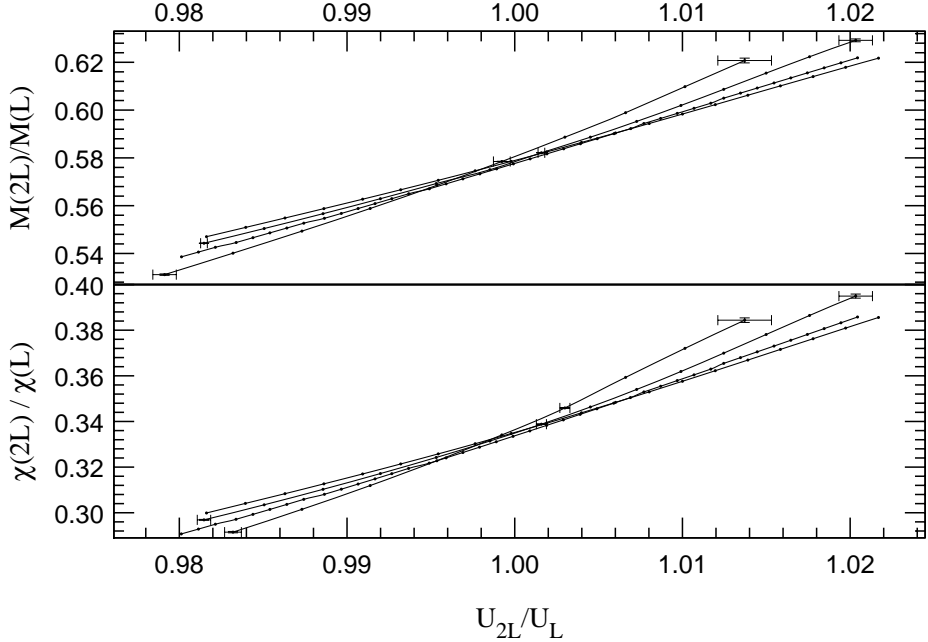}
\caption{Quotients to obtain $\beta/\nu$ and $\gamma/\nu$.} 
\label{l2l.ps}
\end{figure}

For large $L$  
in the critical region we should obtain a single curve,
the deviations corresponding to corrections to scaling.
In the lower part of Figure 7 we plot the same function for
susceptibility.

The values for $\gamma$ and $\beta$ are summarized in Table \ref{TABLE_EXPO}.

\begin{table}[h]
{
\begin{center}
{
\begin{tabular}{|c|c|c|c|}\hline
Lattice sizes  &$\gamma / \nu$  &$\beta / \nu$   &$\nu$      \\  \hline
\multicolumn{4}{|c|}{(without logarithmic corrections)} \\ \hline
$6/12$                    &2.417(3)    &0.791(4)  &0.474(10) \\  \hline
$8/16$                    &2.403(3)    &0.792(6)  &0.483(8)  \\  \hline
$10/20$                   &2.410(2)    &0.790(4)  &0.471(6)  \\  \hline
$12/24$                   &2.403(5)    &0.797(5)  &0.483(7)  \\  \hline
$20/24$                   &2.398(5)    &0.802(4)  &0.487(6)  \\  \hline
\multicolumn{4}{|c|}{(with logarithmic corrections)} \\ \hline
$6/12$                    &2.301(3)    &0.849(4)    &0.484(9)   \\    \hline
$8/16$                    &2.300(3)    &0.850(5)    &0.489(7)   \\   \hline
$10/20$                   &2.315(2)    &0.843(3)    &0.488(5)   \\   \hline
$12/24$                   &2.314(2)    &0.842(5)    &0.487(5)     \\   \hline
$20/24$                   &2.317(5)    &0.839(4)    &0.498(5)  \\   \hline
\end{tabular}
}
\end{center}
}
\caption[a]{ Critical Exponents for the PM-PAF phase transition in the 
$\rm{F}_4$ lattice}
\protect\label{TABLE_EXPO}

\end{table}

\subsection{Logarithmic corrections}

 We now address the question of the possibility of logarithmic corrections
in the AF O(4) model. For the thermal critical exponents, the situation
seems clear, they are compatible with the classical exponent $\nu = 0.5$.
For the magnetic exponents, the situation is more involved. In principle,
one can think that they disagree from MFT due to logarithmic corrections.
We have no perturbative predictions about the form in which these
corrections would affect $\xi_L$ for the AF case. However, one
expects that such corrections slightly modify the critical exponents,
as occurs in the FM case.
It could be possible that logarithmic corrections modify largely the
previous critical exponents and drift them to the FM ones. To sort this
out, we have considered the possibility of
a behavior FM like, so that $\xi_L \sim L(\ln L)^{1/4}$. 
In the lower part of Table \ref{TABLE_EXPO} we quote the values of the
critical exponents for the PAF phase transition when logarithmic
corrections are included (formula (\ref{QUOLOG})). We see how in effect
the magnetic critical exponents are too far from the classical ones
for being the result of a logarithmic correction to the MFT predictions.
It is interesting to compare this situation with that in the RP$^2$
model in $d=4$ \cite{RP24} where small deviations from MFT exponents
can be explained as logarithmic corrections.

\section{Conclusions and outlook}

We have studied the phase diagram of the four dimensional O(4) 
model with first and second neighbors couplings. For $\beta_2 <0$
we find a region non-trivially related with the FM one,
in which the system is AF ordered in some plane.
The phase  transition between the disordered region and this
PAF region seems to be second order.

We also compute the critical exponents on this line at $\beta_1=0$
($\rm{F}_4$ lattice) by means of FSS techniques. We found that up to
$L=24$ the exponents are in disagreement with the Mean Field
predictions.  Specifically, from our $\gamma/\nu$ estimation (or
$\beta/\nu$ using hyper-scaling relation) the exponent $\eta$
associated with the anomalous dimension of the field is $\eta\approx-0.4$.
This fact itself would imply the non-triviality of the theory because
Green functions would not factorize anymore.  One cannot discard that
the observed behavior were transitory. However, the stability of our
measure of $\gamma/\nu$ for lattice sizes ranging from $L=6$ to
$L=24$, which are more than a hundred of standard deviations apart
the MF value, makes very unlikely this hypothesis. Actually, it would
be possible to obtain triviality also with a logarithmic exponent in equation 
(\ref{QUOLOG}) different from 1/4. We can fix the critical exponents
to its MF value and compute this parameter from the numerical data. The
results obtained shown a non asymptotic behavior, with values ranging
from 0.8 to 1.2 for the lattices used. A logarithmic fit is not
satisfactory because of the non-asymptoticity and the large value of
the logarithmic exponent, but larger lattices sizes
are needed in order to get a more conclusive answer.

The behavior of the specific heat does not show any first order
signature, but we have been not able to obtain a reliable estimation
of the $\alpha$ exponent.

We have also measured the Binder cumulant at the critical point,
finding that it stays almost constant when increasing the lattice
size.  If this is not a transient effect, and logarithmic corrections
are finally ruled out, it would correspond a
nonzero value of the renormalized constant in the thermodynamical
limit.

\section{Acknowledgments}

Authors are grateful to J.L. Alonso, H.G. Ballesteros,
V. Mart\'{\i}n-Mayor, J.J. Ruiz-Lorenzo, C.L. Ullod 
and D. I\~niguez for helpful discussions
and comments. We thank specially J.M. Carmona for useful discussions
concerning the logarithmic corrections.
 One of us (I.C.) wishes to thank R.D. Kenway his kind
hospitality at the Department of Physics and Astronomy, Univ. of
Edinburgh, as well as to Edinburgh Parallel Computing Center (EPCC)
for computing facilities where part of these computations have been
done under the financial support of TRACS-EC program.  This work is
partially supported by CICyT AEN94-0218, and
AEN95-1284-E. I.C. holds a
Fellow from Ministerio de Educaci\'on y Ciencia.


\begin{thebibliography}{99}

\bibitem{BRY}
D. Brydges, J. Frohlich and T. Spencer
{\sl Commun. Math Phys.} {\bf 83, p 123} (1982)
\bibitem{AIZ}
M. Aizenman
{\sl Phys. Rev. Lett.} {\bf 97, p 1} (1981)
\bibitem{FRO}
J. Frolich, 
{\sl Nuc. Phys.} {\bf B200 [FS4], p 281} (1982)
\bibitem{LOG}
E. Brezin, J.C. Le Guillou and J. Zinn-Justin. 
{\sl ``Field Theoretical approach to critical phenomena''} in 
{\it Phase Transitions and Critical Phenomena} {ed. C. Domb and
M.S. Green (Academic Press, London)} (1976).
\bibitem{WEG}
F.J. Wegner and E.K. Riedel
{\sl Phys. Rev.}{\bf B7, p 248} (1973)
\bibitem{LUS}
M. Luscher and P. Weisz
{\sl Nuc. Phys.} {\bf B299 [FS20], p 25} (1987)
\bibitem{CALL}
D. E. Callaway, 
{\sl Nuc. Phys.} {\bf B233, p 189} (1984)
\bibitem{U1H}
RTN Collaboration: J.L. Alonso et al.,  
{\sl Nuc. Phys.} {\bf B405, p 574} (1993)
\bibitem{LANG}
R. Kenna and C.B. Lang
{\sl Nuc. Phys.} {\bf B393, p 461} (1993)
\bibitem{LAT}
R. Kenna and C.B. Lang
{\sl Nuc. Phys.} {\bf B30 p 697} (Proc. Suppl.) (1993)
\bibitem{MONT}
I. Montvay
{\sl Nuc. Phys.} {\bf B26 p 57} (Proc. Suppl.) (1991)
\bibitem{RP2}
H. G. Ballesteros, L.A. Fern\'andez, V. Mart\'{\i}n-Mayor and
A. Mu\~noz-Sudupe, 
{\sl Phys. Lett.} {\bf B378, p 207} (1996);
{\sl hep-lat/9605037}, to appear in  {\sl Nuc. Phys. {\bf B}}    
\bibitem{RP22}
E. Seiler and K. Yildirim
{\sl hep-lat E-preprint 9609030}
\bibitem{RP24}
H.G. Ballesteros, J.M. Carmona, L.A. Fern\'andez, V. Mart\'{\i}n-Mayor
A.~Mu\~noz Sudupe and A.~Taranc\'on.
{\sl preprint DFTUZ 96/20}
\bibitem {PLUM}
M.L. Plumer and A. Caill\'e, J. Appl. Phys. {\bf 70, p 5961} (1991).
\bibitem {KAWA}
H. Kawamura, J. Phys. Soc. Jpn. {\bf 61, p 1299} (1992).
\bibitem {MORI}
H. Mori and M. Hamada, Physica B {\bf 194-196, p 1445} (1994).
\bibitem {ISING2d}
J.L. Mor\'an-L\'opez, F. Aguilera-Granja and J.M. S\'anchez,
J.Phys.: Cond. Matt. {\bf 6, p 9759} (1994).
\bibitem{PAJJ}
G. Parisi and J. J. Ruiz-Lorenzo,
J. Phys. A (Math and Gen.) {\bf 28, p L395} (1995).
cond-mat preprint 9503016. To appear in J. Phys. A.
\bibitem{ISING}
J.L. Alonso, J.M. Carmona, J. Clemente, L.A. Fern\'andez,
D. I\~niguez, A. Taranc\'on and C.L. Ullod, 
{\sl Phys. Lett.} {\bf B376, p 148} (1996)
\bibitem{BOCK}
W. Bock, A.K. De, K. Jansen, J. Jersak, T. Neuhaus and J. Smit
{\sl Nuc. Phys.} {\bf B344, p 207} (1990)
\bibitem{BOCK2}
W. Bock, A.K. De, C. Frick, J. Jersak and T. Trapenberg
{\sl Nuc. Phys.} {\bf B378, p 652} (1992)
\bibitem{EBI}
T. Ebihara and K. Kondo
{\sl Prog. Theor. Phys.} {\bf vol 87, 4} (1992)
\bibitem{JL}
J.L. Alonso, Ph. Boucaud, F. Lesmes and E. Rivas.
{\sl Phys. Lett.} {\bf B329, p 75} (1994)
\bibitem{GALLA}
G. Gallavotti and V. Rivasseau, 
{\sl Phys. Lett.} {\bf B122, p 268} (1983)
\bibitem{POLO}
Jochen Fingberg and Janos Polonyi, 
{\sl hep-lat preprint 9602003}                                     
\bibitem{FISH}
P.R.Gerber and M.E. Fisher,
{\sl Phys. Rev.} {\bf B10, p 4697} (1974)
\bibitem{BINDER}
K. Binder, 
{\sl Z. Phys.} {\bf B43, p 119} (1981)
\bibitem{BREZIN}
E. Brezin and J. Zinn-Justin
{\sl Nuc. Phys.} {\bf B257 [FS14], p 867} (1985)
\bibitem{PAR}
Giorgio Parisi,
{\it Statistical Field Theory}, {\sl Ed. Addison Wesley} (1992).
\bibitem{BRE}
E. Brezin
{\sl J. Phys.} {\bf 43, p 15} (1982)
\bibitem{ITZ}
C. Itzykson and J.M.Drouffe, 
{\it Statistical Field Theory}, {\bf Vol. 1}, 
{\sl Cambridge University Press} (1989)
\bibitem{CARDY}
{\it Finite size scaling, Current Physics - Sources
and comments},  
{\sl vol. 2, Ed. J.L. Cardy}, {\sl Elsever Science Publishers
B. V.} (1988)
\bibitem{FS}
A.M. Ferrenberg and R.H. Swendsen, 
{\sl Phys. Rev. Lett.} {\bf B61, p 2635} (1988)
\bibitem{BELLAC}
Michel Le Bellac
{\it Des phenomenes critiques aux champs de jauge} 
{\sl Inter-Editions / Editions du CNRS, Paris} (1988)
\bibitem{ALLES}
B. Alles, A. Buonanno and G. Cella
{\sl hep-lat E-preprint 9609030} (To appear in the Lattice '96 Proc.
Suppl. Nuc. Phys. B)
\bibitem{SEILER}
E. Seiler and A. Patrasciou
{\sl hep-lat E-preprint 9608138}
\bibitem{BHA2}
G. Bhanot
{\sl Nuc. Phys.} {\bf B17, p 653} (Proc. Suppl.) (1990)
\bibitem{POTTS} 
L.A. Fern\'andez, M.P. Lombardo, J.J Ruiz-Lorenzo and A. Taranc\'on, 
{\sl Phys. Lett.} {\bf B277, p 485}(1992)


\end{thebibliography}
\end{document}